\documentclass[twocolumn,showpacs,nofootinbib]{revtex4}

\include{amssymbol}
\input{epsf}
\usepackage{color}

\def\AJ{{\it Ap. J.} }
\def\AJL{{\it Ap. J. Lett.} }

\def\ANJ{{\it Astron. J.} } 
\def\AP{{\it Ann. Phys.} }

\def\ASAS{{\it Astron. and Astrophys.} }

\def\BAPS{{\it Bull. Am. Phys. Soc.} }

\def\CQG{{\it Class. Quantum Gravity} }

\def\GRG{{\it Gen. Relativity and Gravitation} }

\def\IJMP{{\it Int. J. Mod. Phys.} }

\def\MNRAS{{\it Mon. Not. R. Ast. Soc.} }

\def\NC{{\it Il Nuovo Cimento} }

\def\PL{{\it Phys. Lett.} }
\def\PR{{\it Phys. Rev.} }
\def\PRL{{\it Phys. Rev. Lett.} }
\def\PRTS{{\it Physics Reports} }

\def\RMP{{\it Rev. Mod. Phys.} }

\def\al{\alpha} \def\be{\beta} \def\ga{\gamma} \def\de{\delta}
\def\ep{\epsilon}   
\def\th{\theta}   \def\ka{\kappa}
   
\def\si{\sigma}   
   
\def\om{\omega} \def\Ga{\Gamma} \def\De{\Delta} 
 \def\Si{\Sigma}  
  \def\mn{{\mu\nu}} \def\cl{{\cal L}}

 \def\frac#1#2{{\textstyle{{#1}\over
{#2}}}} 
\def\lsim{\mathrel{\rlap{\lower4pt\hbox{\hskip1pt$\sim$}}
\raise1pt\hbox{$<$}}}
\def\gsim{\mathrel{\rlap{\lower4pt\hbox{\hskip1pt$\sim$}}
\raise1pt\hbox{$>$}}} \def\sqr#1#2{{\vcenter{\vbox{\hrule height.#2pt
\hbox{\vrule width.#2pt height#1pt \kern#1pt \vrule width.#2pt} \hrule
height.#2pt}}}}
\def\square{\mathchoice\sqr66\sqr66\sqr{2.1}3\sqr{1.5}3}
\def\beq{\begin{equation}} \def\eeq{\end{equation}}
\def\beqa{\begin{eqnarray}} \def\eeqa{\end{eqnarray}}
\def\eq#1{Eq. (\ref{#1})}

\begin{document}

\title{Gravitational collapse in non-minimally coupled gravity:\\finite density singularities and the breaking of the no-hair theorem}

\author{Jorge P\'aramos}
\homepage{web.ist.utl.pt/jorge.paramos}
\email{paramos@ist.edu}

\author{Catarina Bastos}
\email{catarina.bastos@ist.utl.pt}

\affiliation{Instituto de Plasmas e Fus\~ao Nuclear, Instituto Superior T\'ecnico, Universidade T\'ecnica de Lisboa\\Av. Rovisco Pais 1, 1049-001 Lisboa, Portugal}

\date{\today}

\begin{abstract}

In this work we study the dynamics of gravitational collapse of a homogeneous dust sphere in a model exhibiting a linear non-minimal coupling between matter and curvature. The evolution of the scale factor and the matter density is obtained for different choices of the Lagrangian density of matter, highlighting the direct physical relevance of the latter in this theory. Following a discussion of the junction conditions and boundary terms in the action functional, the matching with the outer metric and event horizon are analyzed.

We find that a distinct phenomenology arises when compared with standard results for the Oppenheimer-Snyder collapse, namely the possibility of finite density black holes and the breaking of the no-hair theorem, due to a dependence of the end state of a black hole on the initial radius of the spherical body.

\end{abstract}


\maketitle 

\section{Introduction}

Two of the major challenges faced by contemporary cosmology are the nature of dark energy \cite{quintessence} (see Ref. \cite{Copeland} for a review) and dark matter \cite{bertone} (or perhaps a unification of the two \cite{Chaplygin}), which account for $\sim 96\%$ of the Universe, and the search for a more encompassing theory of gravity. A rather straightforward approach for the latter problem resorts to the substitution of the linear scalar curvature term in the Einstein-Hilbert action with a function of the scalar curvature, $f(R)$ \cite{f(R)} or other scalar invariants of the theory: extensions relying on a functional dependence of the action on the Gauss-Bonnet invariant $G= R^2 - 4R_\mn R^\mn + R_{\al\be\mn} R^{\al\be\mn}$ \cite{GB} are the most well studied theories, given their invoked origin in a low-energy effective description of String Theory \cite{GBstring} and strong implications in braneworld scenarios \cite{GBbranes}.

The  more tractable class of $f(R)$ models has had considerable success in replicating the early period of rapid expansion of the universe, as shown by the Starobinsky inflationary model $f(R)=R + \al R^2$ \cite{Staro}. At late times, the accelerated expansion of the Universe has also been addressed suitably \cite{fRexp}. Solar System tests, mostly arising from the parameterized post-Newtonian (PPN) metric coefficients derived from this extension of General Relativity (GR), have also been discussed \cite{PPN}. A clear phenomenological consequence of $f(R)$ theories is the addition of an increasing, repulsive contribution to the Newtonian potential, for power law terms $f(R) = f_0 R^n$ \cite{flat}. Aside from the more usual metric affine connection (where the affine connection is taken {\it a priori} as depending on the metric), the so-called Palatini approach \cite{Palatini} (where both the metric and the affine connection are taken as independent variables) has also been considered.
 
Further expanding on this elegant generalization of GR, another interesting possibility has arisen: that the coupling between matter and geometry in the Einstein-Hilbert action is non-minimal --- {\it i.e.} not enforced solely by the invariant $\sqrt{-g} d^4x$ and the use of the metric to raise and lower indexes and the associated covariant derivative. A non-minimal coupling would imply that geometric quantities (such as the scalar invariants) could explicitly appear in the action \cite{Lobo} (see also Ref. \cite{Goenner} for an early proposal in the context of Riemann-Cartan geometry). This leads to the action

\beq S = \int \left[ \ka f_1(R) + f_2(R) \mathcal{L} \right] \sqrt{-g} d^4 x~~. \label{model}\eeq

One is motivated to do so by the presence of a non-minimal coupling stemming from one-loop vacuum-polarization effects in the formulation of Quantum Electrodynamics in a curved spacetime \cite{QED}, as well as in the context of scalar-tensor theories, when considering matter scalar fields \cite{Damour}. Furthermore, it has been shown that a non-minimally coupled theory cannot follow the usual procedure establishing the equivalence between $f(R)$ and a single scalar-tensor theory \cite{scalar1}; indeed, while a theory with two scalar fields may describe the same dynamics as \eq{model}, it still requires one of these fields to appear non-minimally coupled with the matter Lagrangian density \cite{scalar1,CC,Sotiriou1,Sotiriou2,scalar2}.

A non-minimal coupling leads to several phenomenological consequences: in particular, it implies that regions with extreme curvature could lead to considerable deviations from the dynamics predicted by Einstein's theory \cite{Lobo}. A wide range of results has unfolded (see Ref. \cite{review} for a review), including the impact on solar observables \cite{solar}, axisymmetric astrophysical solutions \cite{Martins}, the possibility to account for galactic \cite{DM} and cluster dark matter \cite{cluster}, a mechanism for mimicking a Cosmological Constant at astrophysical scales \cite{CC}, post-inflationary reheating \cite{reheating} or the current accelerated expansion of the universe \cite{cosmo} (also including the so-called ``extended quintessence'' \cite{extquintessence}). Finally, a thorough discussion of the relevance of the choice of Lagrangian density for a perfect fluid and its direct impact on the dynamics of a non-minimally coupled theory was discussed in Ref. \cite{fluid}. This choice will be of the utmost relevance in the current work.

In this study, one addresses how a non-minimal coupling modifies gravitational collapse; similar studies have been performed in the case of standard $f(R)$ theories ({\it i.e.} with $f_1(R) = f(R) \neq R$ and $f_2(R) =1$) \cite{Cembranos,Sharif,Ghosh}. Several simplifications are made, namely that the collapsing body is purely spherical and composed of a homogeneous distribution of dust --- similarly to the well-known Oppenheimer-Snyder (OS) collapse thoroughly studied in GR \cite{OS}. The simplest, linear non-minimal coupling $f_2(R) \sim R$ is considered, and a trivial curvature term $f_1(R) = R$ is taken, so to highlight the effect of the former on the collapse.

\section{Generalities about the Theory}

Variation of \eq{model} with respect to the action yields the modified Einstein field equations,

\beqa \label{field0} && \left( F_1 + {1 \over \ka} F_2 \cl \right) G_\mn =   {1 \over 2\ka} f_2 T_\mn + \\ \nonumber && \tilde{\nabla}_\mn \left( F_1 + {1 \over \ka} F_2 \cl \right) + {1 \over 2} g_\mn \left( f_1 - F_1 R - {1 \over \ka} F_2 R \cl \right)  ~~, \eeqa

\noindent with $\ka = c^4/16\pi G$, $F_i \equiv f_i'(R)$ and $\tilde{\nabla}_\mn \equiv \nabla_\mu \nabla_\nu - g_\mn \square$. The energy-momentum tensor is defined as

\beq T_\mn =-{2 \over \sqrt{-g}}{\de(\sqrt{-g} \cl ) \over \de g^\mn}~~. \label{EMtensor} \eeq

\noindent As expected, GR is recovered from \eq{field0} by setting $f_1(R) = R $ and $f_2(R) = 1$.

The trace of \eq{field0} reads

\beq \left( F_1  + {1 \over \ka} F_2  \cl \right) R  =   {1 \over 2\ka} f_2 T -3 \square \left( F_1 + {1 \over \ka} F_2 \cl \right) + 2 f_1 ~~, \label{trace0}\eeq

\noindent where $T$ is the trace of the energy-momentum tensor.

Resorting to the Bianchi identities, one concludes that the energy-momentum tensor of matter may not be covariantly conserved, as

\beq \nabla_\mu T^\mn={F_2 \over f_2}\left(g^\mn \cl-T^\mn\right)\nabla_\mu R~~. \label{cov} \eeq

\noindent Again, in the absence of a non-minimal coupling, $f_2(R)=1$ and the covariant conservation of the energy-momentum tensor is recovered. This feature implies that the motion of the matter distribution described by a Lagrangian density $\cl$ does not follow a geodesic curve. Clearly, a violation of the Equivalence Principle may emerge if the {\it r.h.s.} of the last equation varies significantly for different matter distributions, which suggests a method of testing the theory and imposing constraints on the associated couplings. This feature is a fundamental characteristic of a non-minimally coupled theory, as shown in Ref. \cite{Sotiriou1}.

\section{Gravitational collapse of a homogeneous fluid}

\subsection{Linear non-minimal coupling}

In Ref. \cite{fluid}, it was argued that the correct Lagrangian density $\cl$ for a perfect fluid is $\cl = -\rho$, as the non-minimal coupling disables the usual on-shell equivalence with other forms (such as $\cl = p$). Notwithstanding, in this study one aims at addressing both forms, so that the impact of choosing a particular Lagrangian density  on gravitational collapse may be gauged directly. 

By considering the usual equation of state (EOS) parameter $\om \equiv p / \rho$ (which may not be a constant), one may encompass both choices by writing $\cl = - \al \rho$, with

\beq \al =
\cases{- \om & , ~~$\cl = p$
\cr 1&,~~  $\cl = -\rho$
} ~~.
 \label{cases}
\eeq

Since one considers a dust distribution, the pressure vanishes and one considers $\om = 0$. For this reason, the parameter $\al$ becomes a binary variable, {\it i.e.} one that adopts only the values $0$ and $1$ (for $\cl = p = 0 $ or $\cl = - \rho$, respectively). A dust distribution implies that there is no supporting pressure to prevent collapse, with no shell crossing during the latter (nor exchange of momentum).

In this work, a linear non-minimal coupling

\beq f_2(R) = 1 + {\ep \over \ka} R~~, \label{linearcoupling} \eeq

\noindent is considered. From the compatibility between a non-minimally coupled preheating mechanism and Starobinsky inflation \cite{reheating}, the dimensionless coupling strength $\ep$ is constrained by

\beq  4.4 \times 10^9 < \ep < 4.4 \times 10^{13} ~~. \label{etarange} \eeq

The energy-momentum tensor for a dust distribution is given by 

\beq T_\mn = \rho u_\mu u_\nu ~~, \label{tensor}\eeq

\noindent where the $4$-velocity $u^\mu$ obeys $u^\mu u_\nu = -1 $ and $ u^\mu \nabla_\nu u_\mu = 0$.

Inserting the form $\cl = - \al \rho$ and Eqs. (\ref{linearcoupling}) and (\ref{tensor}), together with a trivial curvature term $f_1(R) = R$, one finds that \eq{field0} reads

\beqa  \label{field0linear}&& \left( 1 - {\ep \al \over \ka^2} \rho \right) G_\mn =  \\ \nonumber && {1 \over 2\ka} \left( 1 + {\ep \over \ka} R\right) \rho u_\mu u_\nu + {1 \over 2} {\ep \al \over \ka^2} g_\mn R - {\ep\al \over \ka^2} \tilde{\nabla}_\mn  \rho ~~.\eeqa

\noindent Its trace yields

\beq R = { \ka\rho - 3 \ep \al \square \rho \over 2\ka^2 + \ep (2\al - 1 ) \rho } ~~. \label{trace0linear}\eeq
	
The non-conservation of the energy-momentum tensor \eq{cov} becomes

\beq \nabla_\mu T^\mn= - { \ep \over \ka + \ep R }\left( \al g^\mn + u^\mu u^\nu \right) \rho \nabla_\mu R~~. \label{covlinear} \eeq

\subsection{Adaptability of the FRW metric}

In GR, the study of OS collapse is tantamount to the determination of the evolution of the scale factor $a(t)$ appearing in the Friedmann-Robertson-Walker (FRW) metric, as given by the line element

\beq ds^2 = - dt^2 + a^2(t) \left[ {dr^2 \over 1 - kr^2 } + r^2 (d\th^2 + \sin^2\th d\phi^2) \right] ~~, \label{metric} \eeq

\noindent where $k$ is the (negative) spatial curvature.

The use of the above metric implies that one has a position-independent scalar curvature $R$: naturally, this stems from the identification $R = \ka \rho(t)$ that arises from the trace of the usual GR Einstein equations. Naively, one could expect that in the current scenario a homogeneous density also gives rise to a scalar curvature exhibiting only a time dependence, albeit a more convoluted one, as seen in \eq{trace0linear}. This, however, fails to acknowledge that a more evolved metric could give rise to a radial dependence of $R$, via the terms involving the (space dependent) metric components appearing in the D'Alembertian operator $\square$.

To assess this, one therefore adopts a more general metric, given by

\beqa \nonumber && ds^2 = - dt^2 + U(r,t)dr^2 + V(r,t) (d\th^2 + \sin^2\th d\phi^2) ~~, \\ && U(r,t)=A_1^2(t) h(r)~~~~,~~~~V(r,t)=A_2^2(t) r^2~~,\label{metric1}  \eeqa

\noindent so that $t$ measures time according to a in-falling observer (see Ref. \cite{Cembranos} for a similar treatment in $f(R)$ theories). One computes the required differential operator,

\beq \square \rho(t) = -\ddot{\rho} -\left( {\dot{A}_1\over A_1 } +2{ \dot{A}_2  \over A_2} \right) \dot{\rho} ~~, \eeq

\noindent which, by construction, is a function of time only. Thus, from \eq{trace0linear}, it becomes clear that the scalar curvature indeed inherits the homogeneity of the matter density distribution, $R=R(t)$: this condition implies that the general metric \eq{metric1} is indeed reducible to the desired FRW metric \eq{metric} ({\it e.g.} by following the calculations depicted in Ref. \cite{Cembranos}), which will be adopted henceforth.

\section{Evolution of the gravitational collapse}

One now derives the equations of motion driving the dynamics of gravitational collapse. With the assumed FRW metric \eq{metric}, the scalar curvature given by \eq{trace0linear} becomes

\beq \label{curvature} R = { \ka\rho + 3 \ep \al \left( \ddot{\rho} + 3 {\dot{a} \over a } \dot{\rho} \right) \over 2\ka^2 + \ep (2\al - 1 ) \rho } ~~.\eeq

Thus, the temporal component of \eq{covlinear} becomes

\beqa \label{noncons} \dot{\rho} &=&  -\left[ 3 {\dot{a}\over a} + {\ep \over \ka + \ep R} ( 1-\al ) \dot{R} \right] \rho \rightarrow \\ \nonumber \rho(t) &=& \rho_0\left({\ka + \ep R \over \ka + \ep R_0}\right)^{ \al - 1 } \left({a_0 \over a}\right)^3 ~~, \eeqa

\noindent having used $u^\mu = (1,\vec{0})$ in the adopted comoving coordinates, and where the subscript $_0$ denotes initial values (without loss of generality, one sets $a_0\equiv a(t_0) \equiv 1$). Inspection shows that the radial and angular components of \eq{covlinear} vanish trivially.

Substituting \eq{curvature} into the above and considering that $\al = 0,~1$ is a binary variable yields, after some algebra, the closed expression

\beq \label{rho}  \rho(t) = { \rho_0 a^{-3} \over 1 + \ep (1-\al) {\rho_0 \over 2\ka^2} \left( a^{-3} - 1 \right) } ~~. \eeq

\noindent This result is not valid for the case of a non-vanishing pressure, as one cannot freely substitute $-\al $ by the EOS parameter $\omega$, {\it cf.} \eq{cases}: indeed, for $\al \neq 0,~1$, \eq{curvature} and (\ref{noncons}) yield a convoluted non-linear second order equation for $\rho$ (even for a linear EOS $p = \om \rho$ with $\om = {\rm const.}$). As such, the results of this study cannot be straightforwardly generalized for the case of a perfect fluid with pressure.

Even without the knowledge of the evolution of the scale factor $a(t)$, \eq{rho} is revealing of the effect of the adopted non-minimal coupling: if $\cl = -\rho \rightarrow \al=1$, it leads to an unchanged evolution for the increasing density, $\rho \sim a^{-3}$.

If, however, $\cl = p \rightarrow \al=0$, one finds that as the spherical body collapses and $a(t) \rightarrow 0$, it is not infinitely compressed, $\rho \rightarrow \infty$, but rather it tends towards a final state of finite density $\rho_f = 2\ka^2 /\ep \sim M_P^4/\ep$, where $M_P$ is the Planck mass.

Such counter-intuitive result is of course related with the non-conservation of energy, the most striking feature of the non-minimally coupled theory embodied in \eq{model}. Also, notice that the choice of Lagrangian density indeed has a crucial role in the physical outcome of the theory, as discussed in the previous section.

This exotic behaviour is striking: {\it cf.} \eq{etarange}, the coupling strength $\ep \gg 1$ enables a final density well below the Planck scale, $10^{-17} < \rho_f/M_P^4  < 10^{-13}$ --- although a quantum theory of gravity still has to be considered after the spherical body has collapsed to a size below the Planck length. Using $\rho_N = 10^{18}~{\rm kg/m}^3$ as the typical density of a neutron star's core, for comparison, one finds that $  10^{62} < \rho_f/\rho_N < 10^{65} $, so that the final density of the spherical collapse in the $\cl = p \rightarrow \al = 0$ case yields a body compacted many orders of magnitude above the density of atoms: although the assumed perfect fluid is pressureless, the exotic density evolution \eq{rho} may be interpreted as providing an stabilizing effective pressure.

Replacing \eq{curvature} into the modified field Eqs. (\ref{field0linear}) and following some tedious computations, one obtains two dynamical equations of motion relating the density with the scale factor and the spatial curvature,

\beqa \nonumber 0 &=& \left( \ka^2 +2 \ep \al \rho  \right) (k + 2 a \ddot{a} ) +\left( \ka^2 - 4\ep \al \rho  \right) (\dot{a})^2 ~~, \\ \nonumber  {1 \over 6}\ka a^2 \rho &=& \left( \ka^2 - \ep \rho \right) k + \left[ \ka^2 +(3\al-1) \ep \rho \right] (\dot{a})^2 + \\ && \ep (\al-1) a \ddot{a} \rho~~.  \label{fieldset} \eeqa

\noindent Notice that $\al = 0$ leads to the presence of the second derivative of the scale factor in the second equation. 

Evaluating the first one at $t = t_0$, and assuming a collapse with initial null velocity, $\dot{a}(0)=0$, one gets $k=- 2 \ddot{a}_0 > 0 $, regardless of $\al$ (with $\ddot{a}_0 \equiv \ddot{a}(0)$) --- unless $\al = 1$ and $ \rho_0 = -\ka^2 / 2 \ep$ (implying a negative coupling strength $\ep$), in which case $k$ remains undefined. The latter will not be addressed in this study, as a positive coupling strength $\ep$ is assumed.

One may use \eq{fieldset} to eliminate $\ddot{a}$ and write

\beq \label{dotaeq} {1 \over 3}\ka a^2 \rho = \left[2 \ka^2 - (1+\al )\ep \rho \right] k + \left[ 2\ka^2 +(5\al-1) \ep \rho \right] (\dot{a})^2 ~~, \eeq

\noindent which, again assuming $\dot{a}(0)=0$, yields

\beq \label{eqk} k = { 1\over 3\ka} {\rho_0 \over 2 - (1+\al )\ep_0 } ~~,\eeq

\noindent defining the dimensionless parameter

\beq \ep_0 = {\ep \rho_0 \over \ka^2} = 4.8 \times 10^{-76} \ep {\rho_0 \over \rho_N } ~~.\eeq

\noindent Inserting \eq{etarange} and since the spherical body has an initial density $\rho_0 \ll \rho_N$ prior to collapse, one concludes that $\ep_0$ should be extremely small, $\ep_0 \ll 10^{-62} $.

One finds that the non-minimal coupling induces a shift from the value for the spatial curvature found in GR, $k_0= \rho_0/6\ka$. Moreover, a positive spatial curvature also implies that the initial density and the (positive) coupling strength are constrained by

\beq \rho_0 < {\ka^2 \over \ep} {2 \over (1+\al )} ~~.\eeq

Substituting the expression for $\rho(t)$ found in \eq{rho} into the first of Eqs. (\ref{fieldset}) and using the binary identity $\al (\al-1)=0$, one obtains

\beq \label{dota0} 0 = \left( \ka^2 +2 \ep \al \rho_0 a^{-3} \right) (k + 2 a \ddot{a} ) +\left( \ka^2 - 4\ep \al \rho_0 a^{-3} \right) (\dot{a})^2 ~~, \eeq

\noindent leading to a simplified equation of motion for the scale factor 

\beq \label{dota} \dot{a} = - \sqrt{k(1-a) { a^2 +  (a + 1) \al \ep_0 \over a^3 + 2 \al \ep_0}}~~, \eeq

\noindent supplemented by the initial condition $a(0)=1$.

One may easily check that when $a \sim 0$ the scalar curvature behaves as	

\beq \label{curvature2} R \simeq (1+2\al) {3k\over a^2} ~~,\eeq

\noindent so that both choices of Lagrangian density lead to a curvature singularity, $R\rightarrow \infty$ as $a\rightarrow 0$; as discussed before, this is also a density singularity only if $ \al = 1$, since the converse case $ \al = 0$ leads to a point-like object with  finite density $\rho_f$.

Defining the cycloid parameter $\eta$ as,

\beq \label{cycloid} d\eta = {\sqrt{k} \over a} \sqrt{a^3 + a(a+1)\al \ep_0 \over a^3 + 2\al \ep_0} dt ~~,\eeq

\noindent one finds that \eq{dota} becomes the usual evolution equation found in GR,

\beq \label{dotacycloid} a'(\eta) = - \sqrt{ a(1-a)}~~, ~~a(0)=1. \eeq

As in GR, the solution of \eq{dotacycloid} is

\beq \label{solutioncycloid} a(\eta) = {1 + \cos (\eta) \over 2} ~~, \eeq

\noindent so that the gravitational collapse ends when $\eta = \pi$.

In the above, the linear non-minimal coupling manifests itself via the relation between the cycloid time $\eta$ and the coordinate time $t$: if $\cl = p \rightarrow \al = 0$, this amounts only to a shift of the spatial curvature $k$; if $\cl = -\rho \rightarrow \al = 1$, \eq{cycloid} leads to a more complex relation.

\section{Boundary matching}
\label{secjunction}

Assuming that space outside the collapsing spherical body is empty, \eq{field0linear} become the trivial vacuum equations $R_\mn = 0$ found in GR, and as such spacetime is described by a Schwarzschild metric; notice that, if one considers a non-trivial $f_1(R)$, this metric is not the most general solution of \eq{field0}, due to the additional terms present (see Ref. \cite{vacuumfR} for a discussion).

The outer region is endowed with a coordinate system that does not coincide with the one used in the interior region: from spherical symmetry, one sees that only the time and radial coordinates will be different. Labeling these as $t'$ and $r'$, respectively, one writes the outer Schwarzschild metric via the usual line element

\beqa \label{Schwarzschild} ds^2 &=& -\left(1 - {2GM \over r'}\right)dt'^2 + \\ \nonumber && \left(1 - {2GM \over r'}\right)^{-1} dr'^2  + r'^2 (d\th^2 + \sin^2 \th d\phi^2) ~~.\eeqa

One may ask whether the linear non-minimal coupling can lead to change of this outer metric as the collapse ensues, {\it i.e.} of the mass $M$, so that an outer observer would be able to detect it from a variation of  the gravitational potential at a fixed distance $r'$ --- perhaps even towards a final outer Minkowski metric, so that $M \rightarrow 0 $ and the gravitational effect of the spherical body would vanish. The answer is negative: since the scalar curvature $R$ in the vacuum vanishes (as discussed above), the outer Schwarzschild metric remains unchanged throughout collapse (as in GR): the outer mass $M$ is therefore constant and non-vanishing.

For the full spacetime to be well defined, one must ensure that the FRW metric  \eq{metric}, valid inside the collapsing body, matches smoothly with this outer Schwarzschild metric. 

One first requires that the induced metric on the boundary is equal on both sides, as any discontinuities would lead to an ill-defined scalar curvature $R$ --- thus formulating the so-called first junction condition. For this, one first defines the boundary of the collapsing spherical body as a spacelike hypersurface given by the condition $r = r_* = \text{const.}$ (since $r$ is a comoving coordinate) or $r' = R_*(t')$ ({\it i.e.} an external observer sees the boundary receding towards $R_* = 0$).

The tangent vectors are given by

\beq e^\al_a = {\partial y^\al \over \partial x^a}~~,\eeq

\noindent where $y^\al$ are four coordinates used in the inner and outer metrics and $x^a$ ($a= t,\th,\phi$) are the three coordinates parameterizing the boundary \cite{toolkit}. The induced metric $h_{ab}$ on this hypersurface, defined as,

\beq h_{ab} = g_{\al\be} e^\al_a e^\be_b ~~, \eeq

\noindent is then given by the line element

\beqa \label{inducedin} ds^2_{\Si} &=& -dt^2 + a^2 r_*^2 (d\th^2 + \sin^2 \th d\phi^2) = \\ \nonumber &&-\left[ 1 - {2GM \over R_*} - \left( 1 - {2GM\over R_*}\right)^{-1} \left({dr'\over dt'}\right)^2\right] dt'^2 + \\ \nonumber && R_*^2 (d\th^2 + \sin^2 \th d\phi^2)~~. \eeqa
  
By inspection, one obtains the matching conditions for the inner and outer coordinate systems,

\beqa \label{coordinatematching} R_* &=& ar_* ~~, \\ \nonumber \dot{t'} &=& {\sqrt{R_*[R_*-2GM+R_*(\dot{R}_*)^2]} \over R_* - 2GM } ~~, \eeqa

\noindent where the dot still indicates differentiation with respect to $t$.

The smooth crossover between the inner and outer description of spacetime also demands that the derivatives of the corresponding metrics are properly matched. A quick and dirty approach is to require that the derivatives with respect to the radial coordinates are equal; however, this procedure is explicitly coordinate-dependent. A more elegant, coordinate invariant approach relies on the computation of the (dis)continuity conditions for the extrinsic curvature tensor $K_{ab}$ \cite{toolkit}.

In GR, this ensuing second junction condition (in the absence of a thin shell, see Ref. \cite{thin} for a related discussion) translates into the equality $K_{ab}^- = K_{ab}^+$. Erroneously, some assume that this is a universal statement born from geometrical necessity or aesthetic considerations: for example, in Ref. \cite{Sharif}, the unwarranted use of this equality in the context of $f(R)$ theories leads to a time dependent mass $M(t')$ --- thus producing an unphysical outer vacuum with a non-vanishing, time dependent curvature.

Given the above, one concludes that when GR is modified, the second junction condition may in general read $[K_{ab}] \equiv K_{ab}^+ - K_{ab}^- \neq 0 $ \cite{boundarymodified}, with the {\it r.h.s.} reflecting the altered structure of the equations of motion (\ref{field0}).

In order to pursue its correct formulation, two equivalent paths are usually employed, as summarized below. The first works at the level of the equations of motion and resorts to a functional description of the relevant quantities \cite{toolkit,distributions}, thus writing the metric as 

\beq g_\mn = H^- (l) g_\mn^- + H^+ (l) g_\mn^+~~,\eeq

\noindent where $H^\pm$ is the Heaviside step function and $l$ is an affine parameter describing the crossing of the hypersurface at $l=0$. In the interior of the spherical body $l<0$ and $H^- (l) =1$, while outside $H^-(l)=0 $ (and conversely for $H^+$).

The computation of the Ricci tensor and the scalar curvature involve derivatives of the metric; recalling that $H'(l) = \de(l)$, one obtains

\beq R = H^- R^- + H^+ R^+ + \de A~~,\eeq

\noindent where $A = A(g_\mn,g_{\mn,\al},n_\be)$ and $n_\be$ is the unit normal to the hypersurface; the latter enables the relation between the induced and the inner and outer metrics at the boundary,

\beq g^{\al\be} = n^\al n^\be + h^{ab} e^\al_a e^\be_b~~. \label{inducednormalrelation} \eeq

Following the same procedure for the matter content, one writes its energy-momentum tensor as

\beq T_\mn = H^- (l) T_\mn^- + H^+ (l) T_\mn^+~~,\eeq

\noindent with an additional $\de(l)$ term if a boundary layer is present.

By inserting the above expansions for the relevant quantities into \eq{field0} and demanding that terms in $\de(l)$ vanish, one is in principle able to obtain the desired second junction condition. Since many modifications of GR increase the order of the differential operators in the equations of motion, in general this will lead to added terms in $\de$ that yield a discontinuity of the extrinsic curvature across the hypersurface. Also, particular care should be taken with the appearance of crossed terms such as $H^- \de$, as these are ill-defined as functionals --- an issue that could be surpassed with the substitution of the Heaviside step and Dirac delta functions by suitable, convergent, approximations.

The second, much more elegant procedure works at the level of the action of the theory. It relies on the rederivation of the field equations from an action defined within a closed volume of space time: when one follows the usual procedure and applies the Gauss-Stokes theorem, this confinement leads to finite terms (as one can no longer evoke that these vanish at infinity). In order to obtain the same equations of motion, these undesired quantities must be countermanded by a suitable boundary contribution, defined solely on the hypersurface --- the Gibbons-York-Hawking term \cite{GYH} (see Ref.\cite{boundarymodified} for a general derivation in the context of $f(R)$ theories). Finally, by varying the action with respect to the induced metric on the later, one straightforwardly obtains the sought for second junction condition.

In this study, one may quickly conclude as to what method is more profitable: if one adopted the first procedure, the presence of the term $\tilde{\nabla}_\mn \cl$ in \eq{field0linear} would lead to the appearance of derivatives of the Dirac delta, which are functionally defined as $\de'(l) f(l) = - \de f'(l)$. This would lead to extremely cumbersome calculations, thus favoring the second approach outlined above.

With the above in mind, one varies \eq{model} with the adopted forms for $f_1=R$ and $f_2 = 1 + \ep R/\ka$ and a constraint $\de g_\mn = 0$ in the hypersurface $\partial V$. Using

\beq \de R_{\al\be} = \nabla_\si [ g^{\al\be} \de \Ga^\si_{\al\be} - g^{\al\si} \de \Ga_{\al\be}^\be ] ~~, \eeq

\noindent one gets

\beqa && \de S = \int_V \sqrt{-g} d^4 x \times \\ \nonumber && \bigg[  \left( \ka + {\ep \over \ka} \cl \right) R_{\al\be} - {1 \over 2} \ka R g_{\al\be} - {1 \over 2} \left( 1 + \ep{ R \over \ka}\right) T_{\al\be} \\ \nonumber && + \left( \ka + {\ep \over \ka} \cl \right) \nabla_\si [ g^{\al\be} \de \Ga^\si_{\al\be} - g^{\al\si} \de \Ga_{\al\be}^\be ] \bigg] ~~. \eeqa

\noindent The last term, dubbed $\de S_2$, may be integrated via the Gauss-Stokes theorem. For this, one first uses the definition of $\Ga_{\al\be}^\ga$, obtaining

\beqa && \nabla_\si \left(g^{\al\be} \de \Ga^\si_{\al\be} - g^{\al\si} \de \Ga_{\al\be}^\be\right) = \\ \nonumber && g_{\al\be} \square (\de g^{\al\be}) - \nabla_\al \nabla_\be (\de g^{\al\be}) ~~. \eeqa

For convenience, the integral symbols are substituted by an abbreviated notation,

\beqa \int_V X \sqrt{-g} d^4x &\equiv & \left\{X\right\}_V~~,\\ \nonumber \int_{\partial V} X \sqrt{-h} d^3x &\equiv& \left\{X\right\}_{\partial V}~~, \eeqa

\noindent where $h $ is the determinant of the induced metric $h_{ab}$; with this notation, the Gauss-Stokes theorem reads

\beq \left\{ \nabla_\mu X \right\}_V = \left\{ n_\mu X\right\}_{\partial V}~~. \eeq

\noindent Integrating by parts and twice using this theorem, one gets

\beqa && \de S_2 \equiv \left\{ n_\si \left( \ka + {\ep \over \ka} \cl \right) g_{\al\be} \nabla^\si (\de g^{\al\be}) \right\}_{\partial V}  - \\ \nonumber & & \left\{  n_\al \left( \ka + {\ep \over \ka} \cl \right) \nabla_\be (\de g^{\al\be}) \right\}_{\partial V} + \\ \nonumber && \left\{ {\ep \over \ka}(\nabla_\si  \nabla^\si \cl) g_{\al\be} \de g^{\al\be} \right\}_V - \left\{ {\ep \over \ka}( \nabla_\al \nabla_\be \cl) \de g^{\al\be} \right\}_V ~~, \eeqa

\noindent having used $\de g^{\al\be} = 0$ at the boundary.

Thus, the variation of the full action reads

\beqa && \de S = \de \bigg\{ \left( \ka + {\ep \over \ka} \cl \right) R_{\al\be} - {1 \over 2} \ka R g_{\al\be} - \\ \nonumber && {1 \over 2} \left( 1 + \ep{ R \over \ka}\right) T_{\al\be} - {\ep \over \ka} \De_{\al\be} \cl \bigg\}_V + \\ \nonumber && \de \left\{ \left( \ka + {\ep \over \ka} \cl \right) \left[ n_\si g_{\al\be} \nabla^\si (\de g^{\al\be}) - n_\al \nabla_\be (\de g^{\al\be}) \right] \right\}_{\partial V}~~. \eeqa

\noindent The first two lines yield the modified field \eq{field0linear}; the last line must be balanced by an adequate boundary term, given by

\beq \de S_b = \de \left\{ \left( \ka + {\ep \over \ka} \cl \right) \left[  n_\al \partial_\be (\de g^{\al\be}) - n_\si g_{\al\be} \partial^\si (\de g^{\al\be}) \right] \right\}_{\partial V}   ~~, \eeq

\noindent since on the boundary $\de g^{\al\be} = 0$ and thus $ \nabla_\si (\de g^{\al\be}) = \partial_\si (\de g^{\al\be}) $.

Since $\de g^{\al\be}$ does not vary in the hypersurface, its derivative along the tangent vectors vanishes,

\beq e^\al_a e^\be_b h^{ab} \partial_{\be} (\de g_{\si\al}) = 0 ~~,\eeq

\noindent so that, from \eq{inducednormalrelation}, one obtains the simplification

\beqa && (g^{\al\be} - n^\al n^\be ) \partial_{\be} (\de g_{\si\al}) = 0 \rightarrow \\ \nonumber && \partial_\al (\de g^{\si\al}) =  n_\al n^\be \partial_{\be} (\de g^{\si\al}) ~~, \eeqa

\noindent implying that

\beqa && n_\al \partial_\be (\de g^{\al\be}) - n_\si g_{\al\be} \partial^\si (\de g^{\al\be}) = \\ \nonumber && n_\si (n_\al n_\be - g_{\al\be}) \partial^{\si} (\de g^{\al\be}) = \\ \nonumber && -n^\si e_\al^a e_\be^b h_{ab} \partial_\si (\de g^{\al\be}) ~~. \eeqa

As the variation of the trace of the extrinsic curvature $K = h^{ab} K_{ab}$ with respect to $g^{\al\be}$ is given by 

\beq \de K = {1 \over 2} n^\si e_\al^a e_\be^b h_{ab} \partial_\si (\de g^{\al\be}) ~~, \eeq

\noindent one gets

\beqa && \de S_b =- 2 \left\{ \left( \ka + {\ep \over \ka} \cl \right) \de K \right\}_{\partial V} ~~. \eeqa

\noindent Since $ 2\de \cl = ( g_{\al\be} \cl - T_{\al\be} )\de g^{\al\be} = 0 $ on the boundary, one finally obtains the full action including boundary terms (written again in standard notation),

\beqa \label{actionboundary} S &=& \int_V \left[ \ka R + \left( 1 + \ep {R \over \ka} \right) \cl \right] \sqrt{-g} d^4 x - \\ \nonumber && - 2 \int_{\partial V} \left[ \left( \ka + {\ep \over \ka} \cl \right) K + \ep \al \ka x(h_{ab},\cl) \right] \sqrt{-h} d^3 x~~. \eeqa

\noindent The term $x(h_{ab},\cl)$ (with dimensions of mass) stems from the fact that one can supplement the boundary terms above with additional contributions involving $h_{ab}$ and $\cl$ only, as these do not show up when variation with respect to $g_{\al\be}$ is performed (again, since $\de \cl = 0$ on the boundary). It is factored by $\ep \al$ as one expects it to be present only when these quantities are non-vanishing ({\it i.e.} one has a non-minimal coupling with $\cl \neq p = 0$).

The second junction condition may be obtained by variation of the above expression with respect to $h^{ab}$ on both sides of the boundary; considering that there is no surface energy-momentum tensor $S_{ab}$ describing a boundary layer, one has 

\beq \label{layer} S_{ab} = -{2 \over \sqrt{-h}}{\de(\sqrt{-h} \cl ) \over \de h^{ab}} = 0 \rightarrow \de \cl = {1\over 2}  \cl h_{ab} \de h^{ab} ~~,\eeq

\noindent so that, after manipulating the tensors, one obtains

\beqa \label{junction2} K^+_{ab} &= & \left(1 - {\ep \al \over \ka^2} \rho \right) K^-_{ab} + {\ep \al \over \ka^2} \rho K^- h_{ab} + \\ \nonumber && + \ep \al \left[ X_{ab} + h_{ab} (x-X) \right] ~~, \eeqa

\noindent defining $X_{ab} \equiv \de x/ \de h^{ab}$ and its trace $X = h^{ab}X_{ab}$.

Computing the extrinsic curvature tensor for the inner and outer metric from its definition $K_{ab}^\pm \equiv e^\mu_a e^\nu_b e^\nabla_\mu n_\nu^\pm$ yields

\beqa \label{extrinsic} K^{-t}_t &=& 0 ~~, \\ \nonumber K^{-\th}_\th =  K^{-\phi}_\phi &=& {\sqrt{1- kr_*^2 } \over ar_* } ~~, \\ \nonumber K^{+t'}_{t'} &=& {GM + R_*^2\ddot{R}_* \over R_* \sqrt{ R_*[R_* - 2GM + R_*(\dot{R})_*^2 ]}} ~~, \\ \nonumber  K^{+\th}_\th = K^{+\phi}_\phi &=& {\sqrt{ R_*[R_* - 2GM + R_*(\dot{R})_*^2] } \over R_*^2} ~~. \eeqa

\noindent One may now use \eq{junction2} to fix the mass $M$, related to the Schwarzschild radius $R_s \equiv 2GM$; the former is expected to differ from the gravitational mass, defined as $M_0 = (4\pi/3) \rho R_*^3$.

\section{Lagrangian density choice $\cl = p$}

The choice of Lagrangian density $\cl = p \rightarrow = 0$ merely leads to a shift in the definition of the cycloid parameter $\eta$ \eq{cycloid} via the spatial curvature $k$, \eq{eqk}: thus, the gravitational collapse is dynamically equivalent to the case of GR, with the fundamental difference that is does not lead to a singularity of infinite density, as seen from \eq{rho}.

Solving for the original time $t$, one obtains

\beq {d t \over d\eta} = {a \over \sqrt{k}} = {1+ \cos \eta \over 2\sqrt{k}} \rightarrow t = {\eta + \sin \eta \over 2\sqrt{k}}~~. \label{etaalpha0} \eeq

\noindent Gravitational collapse ends at a time $\eta_f = \pi$, which translates into

\beq \label{tfalpha0} t_f = {\pi \over 2\sqrt{k}} = {\pi \over 2} \sqrt{{ 6\ka \over \rho_0} \left( 1 - {\ep_0\over 2} \right) }~~.\eeq

\noindent For a positive coupling strength $\ep$, one finds that the final state of finite density $\rho_f$ is attained earlier than in OS collapse, since the spatial curvature is larger. However, given that $\ep_0 \ll 10^{-62}$, this effect is negligible.

\subsection{Apparent and event horizon}

Following the preceding section, if $\cl = p \rightarrow \al = 0$ one expects that the trapped surfaces, apparent and event horizon all occur analogously to OS collapse, with the non-minimal coupling manifesting itself merely through the shifted spatial curvature $k$, \eq{eqk}. Indeed, by introducing the new radial coordinate $\chi = \arcsin \left( \sqrt{k} r\right) $, one sees that the metric \eq{metric} becomes

\beq  \label{metricnew}  ds^2 = {a^2 \over k} \bigg[ - d \eta^2 + d\chi^2 + \sin^2 \chi (d\th^2 + \sin^2\th d\phi^2) \bigg] ~~, \eeq

\noindent so that radial photons follow null geodesics which are straight lines in the $(\eta,\chi)$ plane, as in GR.

The calculation of trapped surfaces and the ensuing apparent and event horizon proceeds accordingly: in particular, the apparent horizon crosses the surface of the star when it has collapsed below the Schwarzschild radius, and becomes fixed at this value.

In order to determine the latter, one resorts to the second junction condition, \eq{junction2}: setting $\al=0$, one sees that the continuity relation $[K_{ab}]=0$ is recovered. In particular, using \eq{extrinsic}, the $tt$ component together with the identification $R_* = ar_*$ yields

\beqa \label{mass0} && 0 = {GM + R_*^2\ddot{R}_* \over R_* \sqrt{ R_*[R_* - 2GM + R_*(\dot{R})_*^2 ]}} \rightarrow \\ \nonumber && R_s \equiv 2GM = - 2a^2 \ddot{a} r_*^3 = k r_*^3 ~~. \eeqa

having used Eqs. (\ref{fieldset}) and (\ref{dota}) with $\al=0$; the same result of course arises from $[K_{\th\th}]=0$. Inserting \eq{eqk}, one gets

\beq M = {4\pi\over 3} {\rho_0 \over 1 - \ep_0/2 }r_*^3 = {M_0 \over 1-\ep_0/2}~~, \eeq

\noindent showing that the mass of the spherical body, as inferred by an outer observer, is increased due to the presence of the non-minimal coupling.

Remarkably, this result shows that a non-minimal coupling can break the no hair theorem: indeed, two stars with the same gravitational mass but different sizes ({\it i.e.} initial densities $\rho_0$) will take a different time to collapse ({\it cf.} \eq{tfalpha0}) and produce black holes with unequal event horizons. This is not unexpected, since several scalar field theories enable black holes with ``hair'' (see Ref. \cite{bekenstein} for a review), and the considered non-minimally coupled theory can be recast as a multi-scalar-tensor theory \cite{scalar2}.

\section{Lagrangian density choice $\cl = -\rho$}

The choice of Lagrangian density $\cl = -\rho \rightarrow \al = 1$ leads to a more convoluted effect of the non-minimal coupling, as the coordinate time $t$ is related to the cycloid parameter $\eta$ through

\beq \label{cycloidalpha1} t = \int_0^\eta {1\over 2\sqrt{k} } \sqrt{ (1+\cos\eta) \left[ (1+\cos\eta)^3 + 16 \ep_0 \right] \over (1+\cos\eta)^2 + 2(3+\cos\eta) \ep_0 } d\eta~~,\eeq

\noindent having substituted \eq{solutioncycloid} into \eq{cycloid}.

\begin{figure} 

\epsfxsize=\columnwidth \epsffile{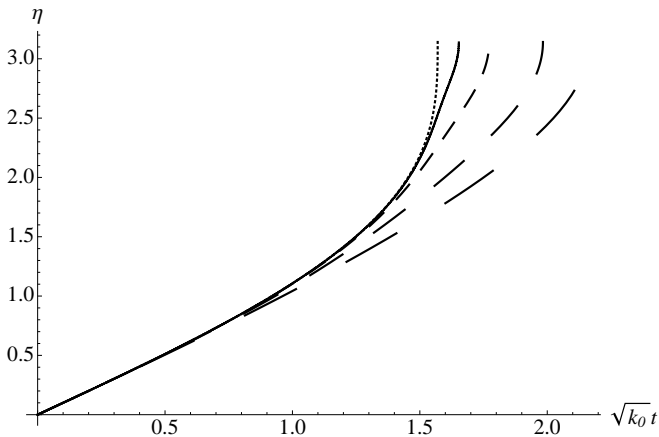}
\caption{Evolution of the cycloid time for different values of $\ep_0 = [10^{-3},~10^{-2},~,10^{-1},~1]$ (full line, small, medium and large dash); dotted indicates $\ep_0 =0$.}
\label{figcycloidalpha1}

\end{figure}

\begin{figure} 

\epsfxsize=\columnwidth \epsffile{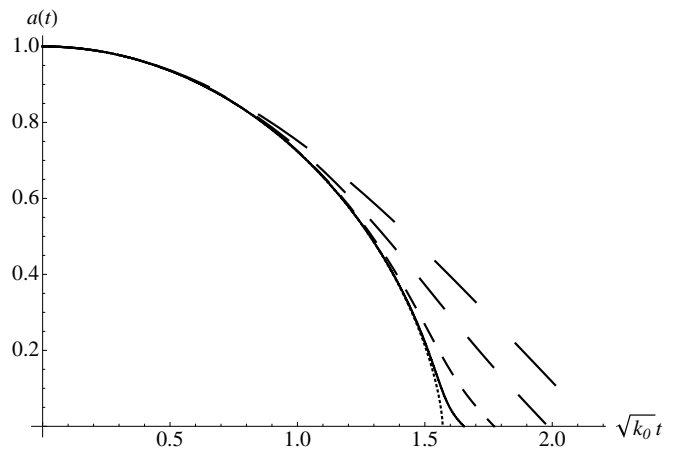}
\caption{Evolution of the scale factor for different values of $\ep_0 = [10^{-3},~10^{-2},~,10^{-1},~1]$ (full line, small, medium and large dash); dotted indicates $\ep_0 =0$.}
\label{figscalealpha1}

\end{figure}

\noindent Since the above cannot be solved analytically, one resorts to a numerical integration, yielding the relation $\eta(t)$ depicted in Fig. \ref{figcycloidalpha1}. Substituting $\eta(t)$ into $a(\eta ) = (1+\cos \eta)/2$ leads to the the modified evolution of the scale factor for different values of $\ep_0$, shown in Fig. \ref{figscalealpha1}. Recall that the constraint on the coupling strength $\ep$ arising from Ref. \cite{reheating}, \eq{etarange}, leads to an extremely small upper bound $\ep_0 \ll 10^{-62}$: for this reason, a larger range of values is plotted, $10^{-3} \leq \ep_0 \leq  1$, to better illustrate the effect of the non-minimal coupling. 
 
One observes that a large deviation of $\eta(t)$ with respect to its GR counterpart arises even if $\ep_0$ is much smaller than unity (where the shift of the spatial curvature \eq{eqk} is negligible), due to the additional term in \eq{cycloidalpha1}.

The relative increase of the collapse time $t_f$ compared to the elapsed period $t_{fOS} = \pi/2\sqrt{k_0}$ for the OS scenario is depicted in Fig. \ref{figtimeend1}, with the former being given by the equality $\eta(t_f) = \pi$.

\subsection{Non-perturbative solution} 

Since $\ep_0 \ll 10^{-62}$, one naturally expects that a perturbative solution of \eq{dota} (with $\al=1$) should ensue. However, this turns out to be unfeasible, as one cannot simply write 

\beq a(t) = a_{OS}(t) + \de a(t)~~, \eeq

\noindent and solve perturbatively for $\de a(t)$; in the above, $a_{OS}(t)$ is the evolution of the scale factor for the Oppenheimer-Snyder collapse of GR ({\it i.e.} with $\ep_0=0$), given by \eq{dota} as

\beq \dot{a}_{OS} = - \sqrt{k {1-a_{OS} \over a_{OS}}}~~. \eeq

\noindent To show this, one first defines $t_{fOS}$ as the end time of OS collapse, $a_{OS}(t_{fOS}) = 0$; since the collapse in the $\al = 1$ non-minimally coupled scenario is delayed, $t_f > t_{fOS}$, when one attains $t \sim t_{fOS}$ the spherical body still has a non-vanishing size and the modification of the scale factor is no longer subdominant, $a(t) \sim \de a (t) \gg a_{OS} (t) \sim 0$, showing that a perturbative solution is disallowed.

\begin{figure} 

\epsfxsize=\columnwidth \epsffile{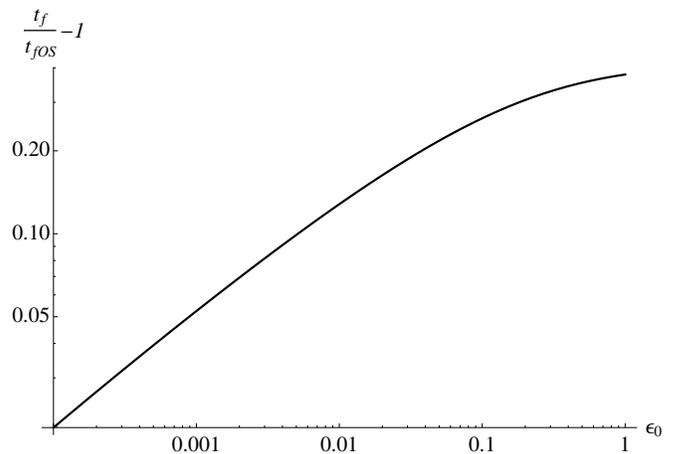}
\caption{Relative increase of the collapse time $ t_e/t_{eOS} -1 $ as a function of $\ep_0$.}
\label{figtimeend1}

\end{figure}

Similarly, one cannot simply expand the integrand of \eq{cycloidalpha1} to first order around $\ep_0 = 0$: doing so produces

\beqa && \nonumber 2\sqrt{k} t \approx \int_0^\eta \left( 1 + \cos \eta + \ep_0 {5 - 4 \cos \eta - \cos^2 \eta \over (1+ \cos \eta)^2} \right)  d\eta \approx \\ & & \ \eta + \sin \eta + \ep_0 {2 \over 3} { (4+\cos\eta)\sin \eta -6\eta \cos^4\left({\eta \over 2}\right)\over (1+\cos \eta)^2 } ~~, \label{cycloidalpha1pert} \eeqa

\noindent naturally yielding a $\ep_0$-dependent correction to the relation $2\sqrt{k} t = \eta + \sin \eta$ found in GR. However, as the size of the spherical body becomes vanishingly small, $\eta_f \rightarrow \pi$, this additional term goes to infinity, $t_f \rightarrow \infty$: contrary to what is shown in Fig. \ref{figtimeend1}, this would signal a never-ending collapse. This again shows that the smallness of $\ep_0 $ does not allow for a perturbative solution.

%%%%%%%%%%%%%%%%%%%%%%%%%%%%%%%%%%%%%%%%%%%%%%%%%%%%%%%%%%%

%%%%%%%%%%%%%%%%%%%%%%%%%%%%%%%%%%%%%%%%%%%%%%%%%%%%%%%%%%%

%%%%%%%%%%%%%%%%%%%%%%%%%%%%%%%%%%%%%%%%%%%%%%%%%%%%%%%%%%%

%%%%%%%%%%%%%%%%%%%%%%%%%%%%%%%%%%%%%%%%%%%%%%%%%%%%%%%%%%%

%%%%%%%%%%%%%%%%%%%%%%%%%%%%%%%%%%%%%%%%%%%%%%%%%%%%%%%%%%%

%%%%%%%%%%%%%%%%%%%%%%%%%%%%%%%%%%%%%%%%%%%%%%%%%%%%%%%%%%%

\subsection{Matching with outer solution}

In the present case $\cl = -\rho$, one unfortunately finds that the inner FRW metric \eq{metric} cannot be suitably embedded with the outer Schwarzschild metric \eq{Schwarzschild}. To ascertain this, one resorts to the junction conditions uncovered before, namely the coordinate matching at the boundary $R_* = a r_*$ and \eq{junction2}, here repeated with $\al=1$:

\beqa \label{junction2alpha1}K^+_{ab} &= & \left(1 - {\ep \over \ka^2} \rho \right) K^-_{ab} + {\ep \over \ka^2} \rho K^- h_{ab} + \\ \nonumber && + \ep \left[ X_{ab} + h_{ab} (x-X) \right] ~~,  \eeqa

By evaluating the above condition using \eq{extrinsic}, one should be able to read the constant $M$. However, this turns out to be unattainable, both for $x = 0$ as well as for a number of candidates for this extra term to the boundary action \eq{actionboundary}, {\it e.g.} $x = \rho$ or $x = h^{ab} \nabla_a \nabla_b \rho$.

Conversely, one may attempt to solve the above for $x$, thus obtaining the additional boundary term that must be considered so that the constant $M$ is recovered: although this is in principle possible, inspection shows it to be extremely cumbersome, with the foreseeable result producing an extremely convoluted and unfounded expression on $\rho$. 

\subsubsection{Comparison with OS collapse}

In the absence of the non-minimal coupling, there is a much more straightforward way to approach the problem, which indeed produces the same results as the painstaking derivation of the Gibbons-York-Hawking boundary action \cite{GYH} and the ensuing second junction condition $[K_{ab}]=0$.

In the standard OS collapse, the absence of pressure indicates that the dust particles on the surface of the spherical body are free-falling along radial geodesics of the outer Schwarzschild metric, so that

\beqa \label{cycloidouter} R_* &=& {R_i \over 2} ( 1 + \cos \eta') ~~, \\ \nonumber \tau &=& \sqrt{R_i^3 \over 8 M} (\eta' + \sin \eta') ~~, \eeqa

\noindent where $\eta'$ is a cycloid parameter related to the proper time $\tau $ of the infalling observer \cite{Wheeler}; the latter is identical with the comoving time of the FRW metric, $\tau = t$.

Recalling the solution \eq{solutioncycloid} and the definition of the inner cycloid parameter \eq{cycloid} (here repeated for convenience),

\beqa a(\eta) &=& {1 + \cos (\eta) \over 2} ~~, \\ \nonumber t &=& \int{a \over \sqrt{k}} \sqrt{a^3 + 2\al \ep_0\over a^3 + a(a+1)\al \ep_0} d\eta ~~,\eeqa

\noindent one finds that the relation $R_* = a r_*$, stemming from the continuity of the induced metric $h_{ab}$ across the boundary ({\it e.g.} the first junction condition) is only valid for all times in the $\al=0$ case (see \eq{etaalpha0}), with $R_i = r_*$ (since $a(0) \equiv 1$) and

\beq \eta = \eta' ~~~~,~~~~ \sqrt{R_i^3 \over 8 GM}= {1 \over 2\sqrt{k}}~~. \eeq 

\noindent The later leads to the result of \eq{mass0}, $2GM = kr_*^3$, valid both for GR (with $M=M_0$) as well as the non-minimally coupled $\al=0$ case.

Following this approach, one traces the impossibility of recovering an expression for $M$ when $\al = 1$ to the mismatch between the definitions of the cycloid parameters \eq{cycloid} and \eq{cycloidouter}. In its turn, this signals a fault in one of the assumptions of the procedure depicted above --- namely that dust particles on the surface of the spherical body free-fall according to radial geodesics of the outer metric.

Indeed, in GR this stems from the condition of vanishing pressure, $p=0$; in the non-minimally coupled scenario, an effective pressure arises, as the $rr$ component of the modified field equations does not vanish, $ p_{eff} \equiv 2\ka g^{rr} G_{rr} \neq 0$, for $\al = 1$ --- and as a result dust particles in the surface experience an additional force that displaces them with respect to radial geodesics. It is null for $\al=0$, so that the above discussion is valid, as attested by the matching between the inner and outer cycloid parameters. 

This is more than a simple mathematical curiosity of the scenario under scrutiny, as it recalls a similar problem in GR: the impossibility of matching the inner and outer spacetimes in the case of a gravitational collapse of a homogeneous sphere $\rho = \rho(t)$ with non-vanishing pressure $ p\neq 0$.

In GR, this can be alleviated by the inclusion of a suitable boundary layer, {\it i.e.} a finite surface energy-momentum tensor $S_{ab}$, as given by \eq{layer}. Such a procedure may also prove helpful in the present context, although it shall not be pursued here: as it stands, the inability to suitably enforce the required matching shows that the gravitational collapse of a linearly minimally coupled homogeneous sphere is well defined only if the Lagrangian density of a perfect fluid is given by $\cl = p$, not $\cl =- \rho$ (see related discussion in Ref. \cite{Sotiriou3}).

\section{Discussion}

In this work we have described the dynamics of gravitational collapse of a dust sphere under the influence of a linear non-minimal coupling, thus extending the familiar Oppenheimer-Snyder collapse. We have examined the different effects that arise due to the choice of Lagrangian density of matter, namely the use of $\cl = - \rho$ or $\cl = p$.

The adopted scenario of a homogeneous spherical body with vanishing pressure is admittedly simplistic, as is the adopted linear form for the non-minimal coupling $f_2(R)$ and the trivial curvature term $f_1(R) = R$; these forms were considered in order to highlight the effect of the former and yield a tractable problem. A generalization of $f_1(R)$ and $f_2(R)$ and the study of a collapsing non-homogeneous sphere with pressure and/or endowed with initial angular momentum and charge should provide for a more evolved phenomenology and yield further insight into the impact of a non-minimal coupling.

Notwithstanding these limitations, the present work shows that a non-minimal coupling can induce significant changes in gravitational collapse. The main results are threefold:

\begin{enumerate}

\item In the $\cl = - \rho$ scenario, the dynamics of gravitational collapse deviates from GR, due to the more evolved dynamics. However, the usual dependence of the density on the scale factor $\rho \sim a^{-3} $ remains, and a point-like singularity with infinite density is attained. Compatibility between a non-minimally coupled preheating mechanism and Starobinsky inflation dictates that the scale factor deviates very weakly from its evolution in GR.

The $\cl = p$ case is much more interesting: although the evolution of the scale factor is qualitatively the same as in OS collapse, the energy-momentum tensor is not conserved. This leads to a modified dependence for the density and, as a result, a geometric point-like singularity ({\it i.e.} where the scalar curvature diverges) is attained with a finite density. Given the largeness of the value of $\ep$, this can fall below the Planckian domain, $\rho \rightarrow \rho_f \ll M_P^4$ (although still many orders of magnitude above the typical density of neutron stars), thus lessening the need for a description of the quantum regime.

\item Analogously to the well-known Gibbons-York-Hawking boundary terms, we have found that an additional contribution to the action functional on the surface of the spherical body must be considered. Its Lagrangian density is of the form $\cl_{\partial V} =  \left( 1 + \ep \cl / \ka^4 \right) K$, with possible, undetermined, additional terms depending on the induced metric and the Lagrangian density of matter.

By varying these boundary terms with respect to the former, we showed that the extrinsic curvature is in general discontinuous across the boundary of the spherical body.

\item The interior description of the gravitational collapse in the $\cl = p$ case is suitably embedded into the surrounding Schwarzschild spacetime via the continuity of the induced metric and the later condition for the extrinsic curvature. This leads to a shift of the mass $M$ of the spherical body (given by the gravitational potential away from it) with respect to the gravitational mass $M_0$: this modification depends not only on the coupling strength $\ep$, but also on the value of the initial density $\rho_0$: as a result, different event horizons arise after collapse for stars with the same initial mass, but distinct radius --- thus breaking the no-hair theorem.

The scenario $\cl = - \rho$ is not so well-behaved: the matching of the inner and outer spacetimes turns out to be unfeasible, unless highly unnatural, apparently arbitrary extra terms are added to the boundary action. From a physical point of view, this can be related to the non-vanishing effective pressure that arises due to the non-minimal coupling --- thus recalling the similar matching problem found in the gravitational collapse of a homogeneous sphere with pressure in GR.

\end{enumerate}

\begin{acknowledgments}

The authors would like to thank O. Bertolami and C. S. Carvalho for fruitful discussions. The authors also acknowledge the referee's valuable remarks and criticism.

The work of CB is sponsored by the {\it Funda\c c\~ao para a Ci\^encia e Tecnologia} (FCT) under the grant $BPD ~23287/2005$. The authors acknowledge the partial support of the FCT project $PDTC/FIS/111362/2009$.

\end{acknowledgments}


\begin{thebibliography}{99}


\bibitem{quintessence}O. Bertolami, \NC {\bf 93 B}, 36 (1986); {\it Fortschr. Physik} {\bf 34}, 829 (1986); S. Perlmutter {\it et al.}, \BAPS {\bf 29}, 1351 (1997); \AJ {\bf 517}, 565 (1999); \AJ {\bf 507}, 46 (1998); R. R. Caldwell, R. Dave and P. J. Steinhardt, \PRL {\bf 80}, 1582 (1998); A. G. Riess {\it et al.}, \ANJ  {\bf 116}, 1009 (1998); B. Schmidt {\it et al.}, \AJ {\bf 507}, 46 (1998); P.M. Garnavich et al., \AJL {\bf 493}, 53 (1998); {\it Science} {\bf 279}, 1298 (1998); \AJ {\bf 509}, 74 (1998); M. Doran {\it et al.}, \AJ {\bf 559}, 501 (2001); C. Rubano and P. Scudellaro, \GRG {\bf 34}, 307 (2002);  S. Capozziello {\it et al.}, \CQG  {\bf 23}, 1205 (2006).

\bibitem{Copeland}E. J. Copeland, M. Sami and S. Tsujikawa, \IJMP {\bf D 15}, 1753 (2006).

\bibitem{bertone}
G. Bertone, D. Hooperand J. Silk, \PRTS {\bf 405}, 279  (2005).

\bibitem{Chaplygin}A. Kamenshchik, U. Moschella and V. Pasquier, \PL {\bf B 511}, 265 (2001); N. Bili\'c, G. Tupper and R. Viollier, \PL {\bf B 535}, 17 (2002); M. C. Bento, O. Bertolami and A. A. Sen, \PR {\bf D 66}, 043507 (2002).

\bibitem{f(R)}S. Capozziello, V. F. Cardone and A. Troisi, \PR {\bf D 71}, 043503 (2005); G. Allemandi, A. Borowiec and M. Francaviglia, \PR {\bf D 70}, 103503 (2004).

\bibitem{GB}D. Lovelock, {\it J. Math. Phys.} {\bf 12}, 498 (1971).

\bibitem{GBstring}M. Gasperini and G. Veneziano, {\it Astroparticle Phys.} {\bf 3}, 317 (1993).

\bibitem{GBbranes}C. Charmousis and J. F. Dufaux, \CQG {\bf 19}, 4671 (2002); J. E. Lidsey and N. J. Nunes, \PR {\bf D 67}, 103510 (2003).

\bibitem{Staro}A. Starobinsky, \PL {\bf B 91}, 99 (1980). 

\bibitem{fRexp}S. Capozziello, \IJMP {\bf D 11}, 483 (2002); S. Capozziello {\it et al.}, \IJMP {\bf D 12}, 1969 (2003);  {\it JCAP} {\bf 08}, 001 (2006); S. Nojiri and S.D. Odintsov, \PR {\bf D 68}, 123512 (2003); \PR {\bf D 74}, 086005 (2006); S. Nojiri, S.D. Odintsov and M. Sami, \PR {\bf D 74}, 046004 (2006); S. M. Carroll {\it et al.}, \PR {\bf D 70}, 043528 (2004); S. M. Carroll {\it et al.}, \PR {\bf D 71}, 063513 (2005); S. Capozziello {\it et al.}, \PL {\bf B 639}, 135 (2006); A. de la Cruz-Dombriz and A. Dobado, \PR {\bf D 74}, 087501 (2006); N. Goheer, J. Larena and P. K. S. Dunsby, \PR {\bf D 80}, 061301 (2009); S. Nojiri {\it et al.}, \GRG {\bf D 42}, 1997 (2010); P. K. S. Dunsby et al., \PR {\bf D 82}, 023519 (2010); P. K. S. Dunsby {\it et al.} {AIP Conf. Proc.} {\bf 1458}, 343 (2011); S. Carloni {\it et al.}, CQG {\bf 29}, 135012 (2012). 

\bibitem{PPN}S. Capozziello, A. Stabile and A. Troisi, gr-qc/0708.0723.

\bibitem{flat}S. Capozziello {\it et al.}, \PL {\bf A 326}, 292 (2004); S. Capozziello, V. F. Cardone and A. Troisi, \MNRAS {\bf 375}, 1423 (2007); J. Mbelek, \ASAS {\bf 424}, 761 (2004).

\bibitem{Palatini}T. Sotiriou, \CQG {\bf 23}, 5117 (2006); N. Poplawski, \CQG {\bf 23}, 2011 (2006); T. Sotiriou and S. Liberati, \AP {\bf 322}, 935 (2007).

\bibitem{Lobo}O. Bertolami {\it et al.}, \PR {\bf D 75}, 104016 (2007).

\bibitem{Goenner}H. F. M. Goenner, {\it Found. Phys.} {\bf 14}, 9 (1984).

\bibitem{QED}I. T. Drummond and S. J. Hathrell, \PR {\bf D 22}, 343 (1980).

\bibitem{Damour}T. Damour and G. Esposito-Far{\`e}se, \CQG {\bf 9} 2093 (1992).

\bibitem{scalar1}P. Teyssandier and P. Tourranc, {\it J. Math. Phys.} {\bf 24}, 2793 (1983); B. Whitt, \PL {\bf B 145}, 176 (1984); J. D. Barrow and S. Cotsakis, \PL {\bf B 214}, 515 (1988); H. Schmidt, \CQG {\bf 7}, 1023 (1990).

\bibitem{CC}O. Bertolami and J. P\'aramos,  \PR {\bf D 84}, 064022 (2011).

\bibitem{Sotiriou1}T. P. Sotiriou and V. Faraoni, \CQG {\bf 25}, 5002 (2008).

\bibitem{Sotiriou2}T. P. Sotiriou and V. Faraoni, \RMP {\bf 82}, 451 (2010).

\bibitem{scalar2}O. Bertolami and J. P\'aramos, \CQG {\bf 25}, 245017 (2008).

\bibitem{review}J. P\'aramos, {\it Proceedings of QSO Astrophysics, Fundamental physics, and Astrometric Cosmology in the Gaia era}, Porto, Portugal, 6 - 9 June 2011; arXiv:1111.2740[gr-qc].

\bibitem{solar}O. Bertolami and J. P\'aramos, \PR {\bf D 77}, 084018 (2008).

\bibitem{Martins} O. Bertolami and A. Martins, \PR {\bf D 85}, 024012 (2012) {[}\href{http://arxiv.org/abs/arXiv:1110.2379}{arXiv:1110.2379}{]}.

\bibitem{DM}O. Bertolami and J. P\'aramos, {\it JCAP} {\bf 1003}, 009 (2010). 

\bibitem{cluster}O. Bertolami, P. Fraz\~ao and J.~P\'aramos, arXiv:1111.3167 [gr-qc], to appear in \PR {\bf D}.

\bibitem{reheating}O. Bertolami, P. Fraz\~ao and J. P\'aramos, \PR {\bf D 83}, 044010 (2011).

\bibitem{cosmo}O. Bertolami, P. Fraz\~ao, and J. P\'aramos, \PR {\bf D 81}, 104046 (2010).

\bibitem{extquintessence}T. Chiba, \PR {\bf D 60}, 083508 (1999); L. Amendola, \PR {\bf D 60}, 043501 (1999).

\bibitem{fluid}O. Bertolami, F. S. N. Lobo and J. P\'aramos, \PR {\bf D 78}, 064036 (2008).

\bibitem{Cembranos}J. A. R. Cembranos, A. de la Cruz-Dombriz and B. M. Nunez, {\it JCAP} {\bf 1204}, 021 (2012).

\bibitem{Sharif}M. Sharif and A. Siddiqa, \GRG {\bf 43}, 73 (2011).

\bibitem{Ghosh}S. G. Ghosh and S. D. Maharaj, \PR {\bf D} 85, 124064 (2012).

\bibitem{OS}J. R. Oppenheimer and H. Snyder, \PR {\bf 56}, 455 (1939).

\bibitem{vacuumfR}S. Capozziello and D. Saez-Gomez, \AP  {\bf 524}, 279 (2012); R. Goswami and G. F. R. Ellis, \GRG {\bf 44}, 2037 (2012).

\bibitem{toolkit}E. Poisson, {\it A Relativist's Toolkit: The Mathematics of Black-Hole Mechanics}, Cambridge University Press (2004).

\bibitem{thin}F. S. N. Lobo and P. Crawford, \CQG {\bf 22}, 1 (2005).

\bibitem{boundarymodified}A. Guarnizo, L. Castaneda and J. M.~Tejeiro, \GRG {\bf 42}, 2713 (2010).

\bibitem{distributions}R. Steinbauer and J. A. Vickers, \CQG {\bf 23}, R91 (2006).

\bibitem{GYH}J. York, \PRL {\bf 28}, 1082 (1972); G. W. Gibbons and S. W. Hawking, \PR {\bf D 15}, 2752 (1977).

\bibitem{bekenstein}J.~D.~Bekenstein,Ê~gr-qc/9808028.

\bibitem{Wheeler}C. W. Misner, K. S. Thorne and J. A. Wheeler, {\it Gravitation}, W. H. Freeman, San Francisco (1973).

\bibitem{Sotiriou3}T. P. Sotiriou, \PL {\bf B 664}, 225 (2008).

\end{thebibliography}
\end{document}